\documentclass[pra,twocolumn,showpacs,superscriptaddress]{revtex4}
\usepackage{amsfonts}
\usepackage{amsmath}
\usepackage{amssymb}
\usepackage{graphicx}

\newcommand{\be}{\begin{eqnarray}}
\newcommand{\ee}{\end{eqnarray}}

\begin{document}

\title{Bi-partite and global entanglement in a many-particle system with
collective spin coupling}
\author{R. G. Unanyan}
\affiliation{Fachbereich Physik, Technische Universit\"at Kaiserslautern, 67653,
Kaiserslautern, Germany}
\affiliation{Institute for Physical Research of Armenian National Academy of Sciences,
Ashtarak-2 378410, Armenia}
\author{C. Ionescu}
\affiliation{Fachbereich Physik, Technische Universit\"at Kaiserslautern, 67653,
Kaiserslautern, Germany}
\affiliation{Institute for Space Science, P.O. Box: MG-23, RO 76911 Bucharest, Romania}
\author{M. Fleischhauer}
\affiliation{Fachbereich Physik, Technische Universit\"at Kaiserslautern, 67653,
Kaiserslautern, Germany}

\begin{abstract}
Bipartite and global entanglement are analyzed for the ground state of a
system of $N$ spin 1/2 particles interacting via a collective spin-spin
coupling described by the Lipkin-Meshkov-Glick (LMG) Hamiltonian. Under
certain conditions which includes the special case of a super-symmetry, the
ground state can be constructed analytically. In the case of an
anti-ferromagnetic coupling and for an even number of particles this state
undergoes a smooth crossover as a function of the continuous anisotropy
parameter $\gamma $ from a separable ($\gamma =\infty $) to a maximally
entangled many-particle state ($\gamma =0$). From the analytic expression
for the ground state, bipartite and global entanglement are calculated. In
the thermodynamic limit a discontinuous change of the scaling behavior of
the bipartite entanglement is found at the isotropy point $\gamma =0$. For $%
\gamma =0$ the entanglement grows logarithmically with the system size with
no upper bound, for $\gamma \neq 0$ it saturates at a level only depending
on $\gamma $. For finite systems with total spin $J=N/2$ the scaling
behavior changes at $\gamma =\gamma _{\mathrm{crit}}=1/J$.
\end{abstract}

\maketitle

%%%%%%%%%%%%%%%%%%%%%%%%%%%%%%%%%%%%%%%%%%%%%%%%%%%%%%%%%%%%%%%%%%%%%%%%%%%%
%%%%%%%%%%%%%%%%%%%%%%%%%%%%%%%%%%%%%%%%%%%%%%%%%%%%%%%%%%%%%%%%%%%%%%%%%%%%

%%%%%%%%%%%%%%%%%%%%%%%%%%%%%%%%%%%%%%%%%%%%%%%%%%%%%%%%%%%%%%%%%%%%%%%%%%%
%%%%%%%%%%%%%%%%%%%%%%%%%%%%%%%%%%%%%%%%%%%%%%%%%%%%%%%%%%%%%%%%%%%%%%%%%%%

%%%%%%%%%%%%%%%%%%%%%%%%%%%%%%%%%%%%%%%%%%%%%%%%%%%%%%%%%%%%%%%%%%%%%%%%%%%

%%%%%%%%%%%%%%%%%%%%%%%%%%%%%%%%%%%%%%%%%%%%%%%%%%%%%%%%%%%%%%%%%%%%%%%%%%%

%%%%%%%%%%%%%%%%%%%%%%%%%%%%%%%%%%%%%%%%%%%%%%%%%%%%%%%%%%%%%%%%%%%%%%%%%%%
%%%%%%%%%%%%%%%%%%%%%%%%%%%%%%%%%%%%%%%%%%%%%%%%%%%%%%%%%%%%%%%%%%%%%%%%%%%
%
%
%
%
%%%%%%%%%%%%%%%%%%%%%%%%%%%%%%%%%%%%%%%%%%%%%%%%%%%%%%%%%%%%%%%%%%%%%%%%%%%
%%%%%%%%%%%%%%%%%%%%%%%%%%%%%%%%%%%%%%%%%%%%%%%%%%%%%%%%%%%%%%%%%%%%%%%%%%%

\section{Introduction}

%%%%%%%%%%%%%%%%%%%%%%%%%%%%%%%%%%%%%%%%%%%%%%%%%%%%%%%%%%%%%%%%%%%%%%%%%%%
%%%%%%%%%%%%%%%%%%%%%%%%%%%%%%%%%%%%%%%%%%%%%%%%%%%%%%%%%%%%%%%%%%%%%%%%%%%

Since the early days of quantum theory it was realized that quantum systems
can possess correlations that do not have a classical counterpart \cite%
{Schroedinger,EPR,Bell}. For a long time this phenomenon, called
entanglement, has been of interest mostly in the context of foundations of
quantum mechanics. With the advent of quantum information science \cite%
{Nielsen-Chuang} it has been realized that entanglement is an essential
resource for efficient computation and communication. This initiated a more
systematic study of its properties. While entanglement in small systems is
by now well understood, many particle entanglement is still a widely open
field. It is well known that quantum correlations and entanglement naturally
occur in interacting many-particle systems but we only begin to understand
their role in the systems \cite{Preskill2000}.

Recently the entanglement properties of quantum systems near the critical
points of quantum phase transitions \cite{Sachdev} have attracted much
attention. Two-particle entanglement has been studied in terms of the
concurrence \cite{Wooters1998} e.g. in one dimensional spin chains \cite%
{OConnor2001,Arnesen2001,Osborne2002,Osterloh2002,Vidal2003}. The
concurrence contains however only limited information about the global
distribution of entanglement, and other measures such as the bi-partite
entanglement between blocks of spins may be of larger interest. Bi-partite
entanglement has been analyzed e.g. in one dimensional quantum spin chains
in \cite{Vidal2003}, where it was shown that it scales logarithmically with
the system size at the critical point with a prefactor determined by the
universality class and saturates in the noncritical regime.

We here study the bi-partite entanglement in a system of spins with a
collective coupling described by a generalization of the
Lipkin-Meshkov-Glick (LMG) Hamiltonian \cite{Lipkin1965}. Since this
Hamiltonian is symmetric under the exchange of particles the Hilbert space
separates into subspaces whose dimensions grow only linearly with the number
of particles, which makes them numerically accessible. In particular we
consider the case of an \textit{even} number of spins and an \textit{%
anti-ferromagnetic} coupling. Furthermore we concentrate on the case where
the Hamiltonian can be factorized in a product of two terms each being
linear in the total spin operators. Under these conditions, which includes
the special case of a super-symmetry (SUSY) \cite{Witten1981}, \cite%
{Cooper1995}, the ground state can be constructed explicitly \cite%
{Unanyan2003}. This ground state undergoes a smooth transition from a
separable to a maximally entangled state as a function of a parameter $%
\gamma $, which characterizes the asymmetry between the collective spin
coupling in $x$ and $y$ directions. The two-particle concurrence in this
system, which due to symmetry is the same for all pair of spins, has been
analyzed in \cite{Vidal2004b}. Very recently also bi-partite entanglement
has been studied in \cite{Stockton2003} and \cite{Vidal-preprint}, however
for the \textit{ferromagnetic} version of the LMG model, where there is a
quantum phase transition. Although in the anti-ferromagnetic case,
considered here, there is no quantum phase transition, we find a
discontinuuous behavior of the entanglement when $\gamma $ is changed.

After discussing the LMG model and its ground state under the condition of
super-symmetry in Sec.II, we analyze the bi-partite entanglement of this
state in terms of the von Neumann entropy \cite{Bennett1996} in Sec.III. We
show that for an isotropic interaction ($\gamma =0$) and in the
thermodynamic limit the entropy grows logarithmically with the spin of the
subsystem with no upper bound. On the other hand for any non vanishing $%
\gamma $ the entropy has an upper limit determined solely by $\gamma $.
Furthermore it becomes a function of the ratio of the subsystem spin to the
total spin rather than a function of the subsystem spin alone. In a finite
system the transition between isotropic and anisotropic behavior occurs at $%
\gamma _{\mathrm{crit}}=J^{-1}$. To understand the saturation of the
entanglement quantitatively, we give in Sec. IV an analytic estimate for the
global entanglement by determining the geometric measure of entanglement.

%%%%%%%%%%%%%%%%%%%%%%%%%%%%%%%%%%%%%%%%%%%%%%%%%%%%%%%%%%%%%%%%%%%%%%%%%%%
%%%%%%%%%%%%%%%%%%%%%%%%%%%%%%%%%%%%%%%%%%%%%%%%%%%%%%%%%%%%%%%%%%%%%%%%%%%

\section{collective spin coupling and super-symmetry}

%%%%%%%%%%%%%%%%%%%%%%%%%%%%%%%%%%%%%%%%%%%%%%%%%%%%%%%%%%%%%%%%%%%%%%%%%%%
%%%%%%%%%%%%%%%%%%%%%%%%%%%%%%%%%%%%%%%%%%%%%%%%%%%%%%%%%%%%%%%%%%%%%%%%%%%

Let us consider an even number $N$ of spin 1/2 particles interacting through
a nonlinear coupling of the collective spin $\hat{J}_{\mu }=\sum_{j=1}^{N}%
\hat{\sigma}_{\mu }^{j}$, where $\hat{\sigma}_{\mu }$ denotes the $\mu $'s
component of the single-particle spin. The interaction is assumed to be of
second order in the total spin and is thus a generalization of the
Lipkin-Meshkov-Glick (LMG) Hamiltonian \cite{Lipkin1965} 
\begin{equation}
H=\alpha \hat{J}_{z}+\beta \hat{J}_{x}^{2}+\hat{J}_{y}^{2}-2\mu \hat{J}_{y}.
\label{hamiltonian}
\end{equation}%
$\alpha $ and $\beta $ are positive real numbers and thus the coupling is of
the anti-ferromagnetic type. $H$ commutes with the total spin $\hat{\mathbf{J%
}}^{2}$ and thus the total Hilbert space separates in sub spaces determined
by the spin quantum number $J$. We here restrict ourselves to the case of
maximum spin, i.e. $J=N/2$. As has been shown in \cite{Unanyan2003}, (\ref%
{hamiltonian}) can be written as a product of two terms linear in the
collective spin operators if $\beta =\alpha ^{2}$: 
\begin{equation}
H=\left( \alpha \hat{J}_{x}+i\hat{J}_{y}-i\mu \right) \left( \alpha \hat{J}%
_{x}-i\hat{J}_{y}+i\mu \right) -\mu ^{2}.  \label{factorizare}
\end{equation}%
$H+\mu ^{2}$ is positive definite, and if $\mu =m$, with $m\in
\{-J,-(J-1),\dots ,(J-1),J\}$, $J=N/2$ being the total angular momentum, it
possesses a non degenerate ground state with $E=0$ obeying 
\begin{equation*}
\left( \alpha \hat{J}_{x}-i\hat{J}_{y}+im\right) \left\vert \Psi
\right\rangle =0.
\end{equation*}%
Since this equation is linear the ground state can be easily constructed,
which yields 
\begin{equation}
|\Psi \rangle =\mathcal{N}(\gamma ,m)\exp [-\gamma \hat{J}%
_{z}]|m_{y}=m\rangle  \label{psi0}
\end{equation}%
where we have introduced the real anisotropy parameter $\gamma $ through $%
\tanh (\gamma )=\alpha \geq 0$. It is interesting to note that an anisotropy
in the spin coupling is reflected here in the non-unitary term $\exp
\{-\gamma \hat{J}_{z}\}$. In the fully isotropic limit $\gamma =0$, the
ground state is the state $m_{y}=m\rangle $ which is entangled for all $%
|m|<J $. In the maximally anisotropic case $\gamma =\infty $, the ground
state is $|m_{z}=-J\rangle $, which is a product state. The loss of
entanglement in this case is due to the non-unitary term $\exp \{-\gamma 
\hat{J}_{z}\}$. Changing $\gamma $, e.g. as function of time, from $\infty $
to $0$ causes a smooth transition from a factorized to an entangled
many-body state.

Due to the symmetry of the coupling all matrix elements of the Hamiltonian
between states corresponding to different total spin $J$ vanish exactly even
for time-dependent parameter. Thus even though the ground state of (\ref%
{hamiltonian}) becomes degenerate for $\gamma=0$ with respect to the total
spin $J$ \cite{Vidal2004b}, the system cannot undergo a first order quantum
phase transition upon changing $\gamma$. In fact the degeneracy in $J$ at $%
\gamma=0$ can easily be lifted by adding a term $-\lambda \hat {\mathbf{J}}%
^2 $ to (\ref{hamiltonian}), which has no effect on $|\Psi\rangle$.

In the following we will restrict ourselves to the most interesting special
case $m=0$. As has been shown in \cite{Wei2003} and \cite{Wei2004} the
collective state $|m=0\rangle $ has the largest global entanglement and
should thus be considered as the state with maximum $N$-particle
entanglement. A generalization to arbitrary $m$ values is rather straight
forward but less instructive. An additional feature of the $m=0$ case is the
presence of a super-symmetry of the LMG Hamiltonian \cite{Unanyan2003}. As a
consequence in every spin sector $J$ the spectrum of (\ref{hamiltonian}) has
for all values of $\gamma $ a nondegenerate ground state and all excited
states are pairwise degenerate \cite{Cooper1995}. As shown in \cite%
{Unanyan2003} the energy gap between the ground state and the pair of first
excited states does not close.

For $m=0$ the ground state (\ref{psi0}) reads explicitly 
\begin{eqnarray}
|\Psi\rangle =\frac{\mathrm{e}^{-\gamma \hat J_z}} {\sqrt{P_{J}(\cosh2\gamma)%
}}|m_y=0\rangle  \label{SUSY-state}
\end{eqnarray}
with $P_{J}$ being Legendre polynomials.

%%%%%%%%%%%%%%%%%%%%%%%%%%%%%%%%%%%%%%%%%%%%%%%%%%%%%%%%%%%%%%%%%%%%%%%%%%%
%%%%%%%%%%%%%%%%%%%%%%%%%%%%%%%%%%%%%%%%%%%%%%%%%%%%%%%%%%%%%%%%%%%%%%%%%%%

\section{Bi-partite entanglement}

%%%%%%%%%%%%%%%%%%%%%%%%%%%%%%%%%%%%%%%%%%%%%%%%%%%%%%%%%%%%%%%%%%%%%%%%%%
%%%%%%%%%%%%%%%%%%%%%%%%%%%%%%%%%%%%%%%%%%%%%%%%%%%%%%%%%%%%%%%%%%%%%%%%%%

In the following section we discuss the entanglement between two arbitrary
partitions of the $N$ particle system in the SUSY ground state of the LMG
model. As mentioned in the introduction it is not important here how the
partitioning is done. Due to the symmetry of the Hamiltonian only the number
of particles in each partition is of relevance.

%%%%%%%%%%%%%%%%%%%%%%%%%%%%%%%%%%%%%%%%%%%%%%%%%%%%%%%%%%%%%%%%%%%%%%%%%%

\subsection{Entropy of entanglement and distribution of Schmidt coefficients}

%%%%%%%%%%%%%%%%%%%%%%%%%%%%%%%%%%%%%%%%%%%%%%%%%%%%%%%%%%%%%%%%%%%%%%%%%%

A generally accepted quantitative measure for the entanglement between two
subsystems $1$ and $2$ if the total system is in a pure state $|\Psi \rangle 
$ is the von Neumann entropy of either of the two subsystems (entropy of
entanglement) 
\begin{equation*}
S(\Psi )=-\mathrm{tr}_{1}\left\{ \rho _{1}\ln \rho _{1}\right\} =-\mathrm{tr}%
_{2}\left\{ \rho _{2}\ln \rho _{2}\right\} ,
\end{equation*}%
where 
\begin{equation*}
\rho _{1,2}=\mathrm{tr}_{2,1}\Bigl\{|\Psi \rangle \langle \Psi
|\Bigr\},\qquad
\end{equation*}%
are the reduced density matrices. $S(\Psi )$ is essentially a measure for
the information loss due to division of the system and ignoring one of the
subsystems. If there is entanglement between $1$ and $2$ in the original
pure state $|\Psi \rangle $ of the total system, the entropy is nonzero. On
the other hand if $|\Psi \rangle $ factorizes there is no information loss
if we ignore one subsystem and the entropy vanishes.

The von Neumann entropy for pure states $S(\Psi )$ is identical to the
minimum relative entropy of entanglement $E_{2}(\Psi )$ \cite{Vedral2002}%
with respect to all bi-partite separable states $\sigma \in \mathcal{S}_{2}$ 
\begin{equation*}
\sigma =\sum_{i}p_{i}\sigma _{1}^{i}\otimes \sigma _{2}^{i}\qquad p_{i}\geq
0,\,\,,\sum_{i}p_{i}=1.
\end{equation*}
\begin{eqnarray}
E_{2}(\Psi ) &=&\underset{\sigma \in \mathcal{S}_{2}}{\min }S\bigl(\Psi
||\sigma \bigr), \\
S\bigl(\Psi ||\sigma \bigr) &=&\text{tr}\left( \rho \log _{2}\rho -\rho \log
_{2}\sigma \right)
\end{eqnarray}%
and $\rho =|\Psi \rangle \langle \Psi |.$

Calculating the von Neumann entropy of a many-particle system is in general
a very nontrivial task due to the exponential growth of the relevant Hilbert
space. We will show now that the von Neumann entropy can be related to the
variance of the distribution of Schmidt coefficients, arranged in an
appropriate order, in the limit of a large number of particles. For the
symmetric spin states considered here this variance can easily be
calculated, which will be done in the following subsection.

Let $\left\vert \Psi\right\rangle $ \ denote a pure state of a quantum
system consisting of two parts labeled $1$ and $2.$ In the case of finite
dimensional spaces, Schmidt's theorem \cite{Peres-book} asserts that any
state $\left\vert \Psi\right\rangle $ in the Hilbert space $H_{1}\otimes
H_{2}$ can be written in the form: 
\begin{equation}
\left\vert \Psi\right\rangle =\sum\limits_{m=1}^{\chi}\lambda_{m}\left\vert
\Phi_{m}^{(1)}\right\rangle \otimes\left\vert \Phi_{m}^{(2)}\right\rangle
\label{Schmidt}
\end{equation}
where $\chi\le \min\{d_{1},d_{2}\},$ $d_{1}$ and $d_{2}$ being the
dimensions of the corresponding Hilbert spaces. $\left\{ \left\vert
\Phi_{m}^{(1)}\right\rangle \right\} $ and $\left\{ \left\vert
\Phi_{n}^{(2)}\right\rangle \right\} $ are sets of orthonormal states for
the subsystems $1$ and $2$, respectively, and $\lambda_{m}$ are the positive
Schmidt coefficients obeying the sum rule 
\begin{equation}
\sum\limits_{m=1}^{\chi}\lambda_{m}^{2}=1.  \label{norma}
\end{equation}
It is easy to see that the entropy of entanglement is related to the Schmidt
coefficients via 
\begin{equation}
S(\Psi)=-\sum\limits_{m=1}^{\chi}\lambda_{m}^{2}\log_{2} \lambda_{m}^{2}.
\label{entropy1}
\end{equation}
The Schmidt rank, i.e. the number of Schmidt coefficients $\chi$ provides a
simple upper bound for the entropy of entanglement $S(\Psi)\le \log_{2}\chi$%
. If $\chi\sim\mathrm{min}\{d_1 , d_2\}$, i.e. if it scales exponentially
with the number of particles, $\log_2\chi$ is a polynomial function of the
number of particles in the smaller of the two subsystems. For the symmetric
coupling considered here, the dimension of the relevant Hilbert space $d$
increases only linear in the number of particles implying a logarithmic
scaling of $\log_2\chi$ with the system size. Thus also the von Neumann
entropy $S$ is expected to scale logarithmically, however with a yet unknown
coefficient. In order to calculate this coefficient it is obviously not
sufficient to use $\log_2\chi$ as an estimate for $S(\Psi)$.

It is possible, however, to find a better estimate for $S(\Psi )$ in terms
of the variance of the distribution of appropriately ordered Schmidt
coefficients. To show this we first note that, as shown in \cite{Beckner1997}%
, the following inequality holds 
\begin{equation}
-\int \left\vert \,f(x)\right\vert ^{2}\log _{2}\,\left\vert
\,f(x)\right\vert ^{2}dx\leq \frac{1}{2}\left( 1+\log _{2}\pi e\right) +\log
_{2}\Delta x  \label{entropyineq}
\end{equation}%
where $\Delta x^{2}=\int (x-\overline{x})^{2}\left\vert f(x)\right\vert
^{2}dx$ is the variance of a probability distribution $f(x)$. This
inequality becomes an equality if $f(x)$ is a Gaussian function.

For large values of $\chi $ the sum in eq.(\ref{entropy1}) can be written as
an integral with $\lambda _{m}^{2}\,\longrightarrow \,\lambda ^{2}(m)$,
representing a continuous, smooth probability distribution. The functional
form of this distribution depends on the ordering of the Schmidt
coefficients $\lambda _{m}$. For symmetric states this ordering can be
chosen in such a way, that $\lambda ^{2}(m)$ becomes to a good approximation
a Gaussian function \cite{Wigner}. In this case the entropy of entanglement
is given by 
\begin{equation}
S(\Psi )=\frac{1}{2}\left( 1+\log _{2}\pi e\right) +\log _{2}\Delta \lambda ,
\label{entropyequal}
\end{equation}%
where $\Delta \lambda $ is the variance of the Schmidt coefficients.

%%%%%%%%%%%%%%%%%%%%%%%%%%%%%%%%%%%%%%%%%%%%%%%%%%%%%%%%%%%%%%%%%%%%%%%%%

\subsection{Clebsch-Gordan decomposition and bi-partite entanglement}

%%%%%%%%%%%%%%%%%%%%%%%%%%%%%%%%%%%%%%%%%%%%%%%%%%%%%%%%%%%%%%%%%%%%%%%%%

We will now calculate the entropy of entanglement using eq.(\ref%
{entropyequal}) by finding a suitable bipartite decomposition of the ground
state (\ref{SUSY-state}). The scaling of $S(\Psi)$ with the system size will
be studied in detail and in particular the prefactor of the logarithm
determined. In the limit of a totally isotropic spin coupling $\gamma=0$ an
explicit analytic expression for the variance of the Schmidt coefficients
and the entanglement can be given. For nonzero values of $\gamma$ numerical
results will be presented.

We start from equation (\ref{SUSY-state}) and make use of the Clebsch-Gordan
decomposition of total angular momentum $J=N/2$ into $J_1$ and $J_2$ with $%
J_1+J_2=J$: 
\begin{eqnarray}
&&\left\vert \Psi\right\rangle =\mathcal{N}(\gamma)e^{-\gamma\hat{J}_{z}
}\left\vert m_{y}=0\right\rangle  \label{psiCG} \\
&&=\mathcal{N}(\gamma)\sum_{m_{1}}\sum_{m_{2}} C_{m_{1} m_{2} m}^{J_{1}\
J_{2}\ J}\, d_{m, 0}^{J}(-i\gamma)\left\vert J_{1}\,m_{1}\right\rangle
\otimes\left\vert J_{2}\,m_{2}\right\rangle  \notag
\end{eqnarray}
Here $C_{m_{1} m_{2} m}^{J_{1} \ J_{2} \ J}$ are the Clebsch-Gordan (Wigner)
coefficients and $m=m_1+m_2$. $d_{m^{^{\prime }}m}^{J}(\beta)$ are the
rotation matrices defined as \cite{Biedenharn1981} 
\begin{eqnarray}
&&d_{m^{\prime}\, m}^{J}\left( \beta\right) =\left\langle J m^{\prime
}\right\vert e^{-i\beta\hat{J}_{y}}\left\vert J m \right\rangle
\label{rotmat} \\
&&\,\, =\sqrt {\frac{(J+m)!(J-m)!}{(J+m^{\prime})!(J-m^{\prime})!}}\left(
\sin\frac{\beta }{2}\right) ^{m-m^{\prime}}\left( \cos\frac{\beta}{2}\right)
^{m+m^{\prime }}  \notag \\
&&\qquad \times\, P_{J-m}^{(m-m^{\prime},\ m+m^{\prime})}(\cos\beta),  \notag
\end{eqnarray}
where $P_{n}^{(\alpha,\beta)}(x)$ are Jacobi polynomials. Since we are
considering the special decomposition for $J=J_{1}+J_{2}$, the Wigner
coefficients have the binomial distribution: 
\begin{equation}
\left( C_{m_{1} m_{2} m}^{J_{1}\ J_{2} \ J}\right) ^{2}=\frac{%
\begin{pmatrix}
2J_{1} \\ 
J_{1}+m_{1}%
\end{pmatrix}%
\begin{pmatrix}
2J_{2} \\ 
J_{2}+m_{2}%
\end{pmatrix}
}{%
\begin{pmatrix}
2J \\ 
J+m%
\end{pmatrix}
}.  \label{CG form}
\end{equation}
This relation combined with (\ref{rotmat}) gives the following decomposition
for the wave function $\left\vert \Psi\right\rangle $: 
\begin{eqnarray}
\left\vert \Psi\right\rangle =\sum_{m_{1}=-J_{1}}^{J_{1}}
\sum_{m_{2}=-J_{2}}^{J_{2}} A_{m_1,m_2}(\gamma)\left\vert
J_{1}\,m_{1}\right\rangle \otimes\left\vert J_{2}\,m_{2}\right\rangle
\label{state-2}
\end{eqnarray}
with 
\begin{eqnarray}
&&A_{m_{1},m_2}(\gamma)= \mathcal{N}(\gamma)\, \, (i)^{m_{1}+m_{2}}\times 
\notag \\
&&\times \sqrt{\frac{(J!)^{2}(2J_{1})!(2J_{2})!}{%
(2J)!(J_{1}+m_{1})!(J_{1}-m_{1})!(J_{2}+m_{2})!(J_{2}-m_{2})!}}  \notag \\
&&\times \left( \coth\frac{\gamma}{2}\right)
^{m_{1}+m_{2}}P_{J}^{(-m_{1}-m_{2},\ m_{1}+m_{2})}(\cosh\gamma).
\label{Schmidt-final}
\end{eqnarray}
(\ref{state-2}) is separable if all coefficients $A_{m_1,m_2}$ factorize.
This is the case in the limit $\gamma\to\infty$, where the $\coth$ term in (%
\ref{Schmidt-final}) approaches unity and the Jacobi polynomials factorize
in $m_1$ and $m_2$.

%********************************************************************

\subsubsection{isotropic spin coupling $\protect\gamma=0$}

%********************************************************************

\begin{figure}[tbh]
\begin{center}
\includegraphics[width= 7 true cm]{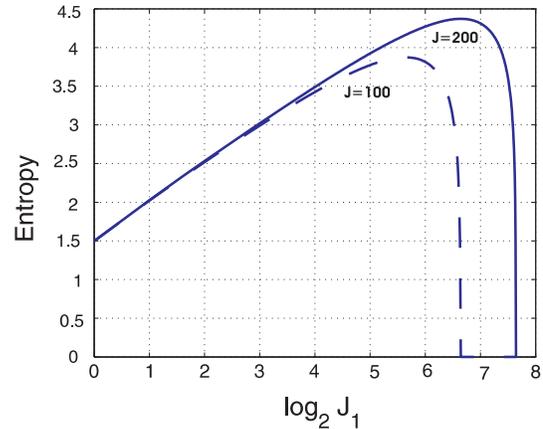}
\end{center}
\caption{Entropy of entanglement for $\protect\gamma =0$ as function of the
logarithm of the subsystem spin $J_{1}$ for different values of the total
spin $J$. Due to the symmetry of $S$ with respect to $J_{1}\leftrightarrow
J-J_{1}$ the curves saturate at $J_{1}=J/2$.}
\label{bipartite-entangle-1}
\end{figure}

Making use of the results of subsection A the summation in (\ref{psiCG}) can
be carried out explicitly for the limit of isotropic spin coupling. In this
case one sees from (\ref{Schmidt-final}) that only coefficients $%
A_{m_{1},m_{2}}$ with $m_{1}+m_{2}=0$ survive. Thus one has the following
decomposition 
\begin{equation*}
\left\vert \Psi \right\rangle =\sum_{m=-J_{1}}^{J_{1}}C_{m-m\ \ \ \
0}^{J_{1}\ J_{2}\ J_{1}+J_{2}}\left\vert J_{1}\,m\right\rangle \otimes
\left\vert J_{2}\,-m\right\rangle
\end{equation*}%
where $J_{2}\geq J_{1}$ was assumed without loss of generality. The Schmidt
coefficients are, therefore, the Clebsch-Gordan coefficients. In the limit $%
J_{2}\gg J_{1}$ , they have a Gaussian form \cite{Wigner} 
\begin{equation*}
\left( C_{m-m\ \ \ \ 0}^{J_{1}\ J_{2}\ J_{1}+J_{2}}\right) ^{2}\approx \frac{%
\exp \left[ -\frac{m^{2}}{J_{1}}\right] }{\sqrt{J_{1}\pi }}.
\end{equation*}%
where $|m|\leq J_{1}\ll J_{2}$. The Gaussian form of the coefficients allows
to make use of relation (\ref{entropyequal}) to calculate the von Neumann
entropy for $J_{1}\ll J$: 
\begin{equation}
S\simeq \frac{1}{2}\log _{2}J_{1}+\frac{1}{2}(1+\log _{2}\pi e),\qquad
\gamma =0  \label{Sgama0}
\end{equation}%
In fig.\ref{bipartite-entangle-1} we have plotted the von Neumann entropy
for $\gamma =0$ as a function of the subsystem spin $J_{1}=N_{1}/2$ for
different values of the total spin $J=N/2$. For $J_{1}\ll J$ a logarithmic
scaling with prefactor 1/2 is evident. When $J_{1}$ approaches $J/2$ the
entropy saturates since $S$ is symmetric with respect to the replacement $%
J_{1}\leftrightarrow J-J_{1}$. It is important to note that for $J_{1}\ll J$
the von Neumann entropy $S$ does not depend on $J$.

%%%%%%%%%%%%%%%%%%%%%%%%%%%%%%%%%%%%%%%%%%%%%%%%%%%%%%%%%%%%%%%%%%%%%%%%%%%%
%%%%%%%%%%%%%%%%%%%%%%%%%%%%%%%%%%%%%%%%%%%%%%%%%%%%%%%%%%%%%%%%%%%%%%%%%%%%%

%************************************************************************

\subsubsection{anisotropic spin coupling $\protect\gamma\ne 0$}

%************************************************************************

\begin{figure}[tbh]
\begin{center}
\includegraphics[width= 8 true cm]{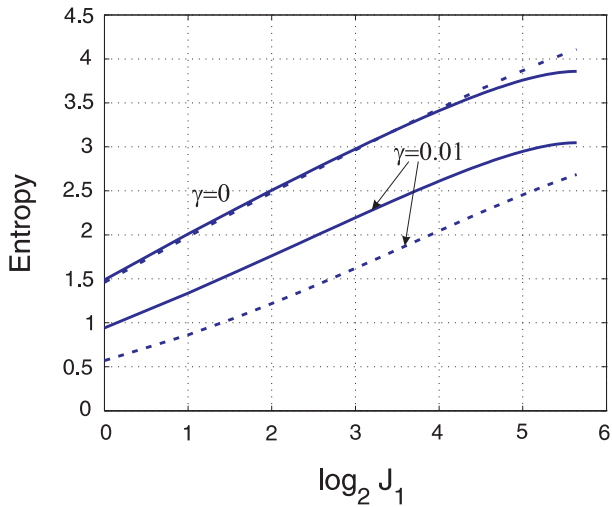}
\end{center}
\caption{Entropy of entanglement for $J=100$ (full line) and $J=200$ (dashed
line) as function of logarithm of subsystem spin $J_{1}$ for different
values of $\protect\gamma $. One recognizes that in contrast to the
isotropic case the entropy now depends on the total spin $J$.}
\label{bipartite-entangle-2}
\end{figure}

\begin{figure}[htb]
\begin{center}
\includegraphics[width= 8 true cm]{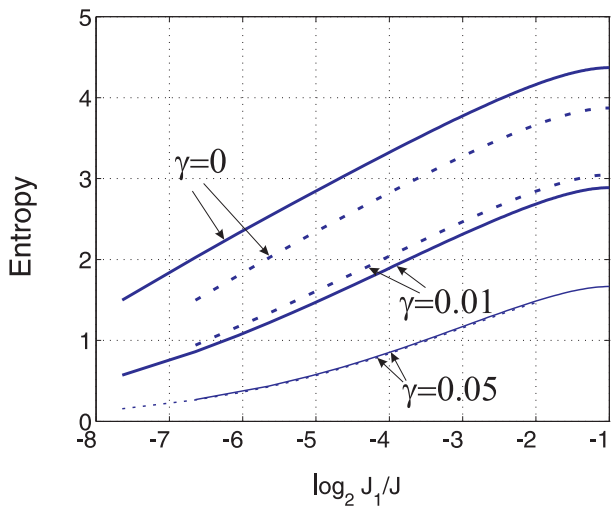}
\end{center}
\caption{Entropy of entanglement for different values of $\protect\gamma$ as
function of $\log_2(J_1/J)$ for $J=100$ (solid line) and $J=200$ (dashed
line). For $\protect\gamma J\ge 1$ the curves become virtually
indistinguishable.}
\label{bipartite-entangle-3}
\end{figure}

If the spin coupling is anisotropic, i.e. if $\gamma \neq 0$, the double sum
in eqs. (\ref{psiCG}) or (\ref{state-2}) remains. Thus in order to discuss
the influence of a finite $\gamma $ it is necessary to explicitly evaluate
the sum in (\ref{entropy1}). We have done this numerically for a total
particle number up to 200 and subsystems up to 100 particles. The results
are shown in figs.\ref{bipartite-entangle-2} and \ref{bipartite-entangle-3}.
As can be seen from fig.\ref{bipartite-entangle-2} in contrast to the
isotropic case $\gamma =0$, the entropy is no longer independent on the
total spin if 
\begin{equation*}
\gamma \geq \gamma _{\mathrm{crit}}\equiv \frac{1}{J}.
\end{equation*}%
In the thermodynamic limit the critical point is $\gamma =0$.

As can be seen from fig.\ref{bipartite-entangle-3} for any $\gamma\ge J^{-1}$
the entropy becomes a function of the logarithm of the fraction of particles 
$J_1/J= N_1/N$.

%%%%%%%%%%%%%%%%%%%%%%%%%%%%%%%%%%%%%%%%%%%%%%%%%%%%%%%%%%%%%%%%%%%%%%%%%%%%
%%%%%%%%%%%%%%%%%%%%%%%%%%%%%%%%%%%%%%%%%%%%%%%%%%%%%%%%%%%%%%%%%%%%%%%%%%%%

%%%%%%%%%%%%%%%%%%%%%%%%%%%%%%%%%%%%%%%%%%%%%%%%%%%%%%%%%%%%%%%%%%%%%%%%%%%%
%%%%%%%%%%%%%%%%%%%%%%%%%%%%%%%%%%%%%%%%%%%%%%%%%%%%%%%%%%%%%%%%%%%%%%%%%%%%%

Our numerical calculations suggest for $J_{1}\ll J$ and $\gamma \gg J^{-1}$
a functional dependence of the form 
\begin{equation*}
S\sim f(\gamma )\log _{2}\Bigl(J_{1}/J\Bigr),\qquad \gamma >\gamma _{\mathrm{%
crit}}
\end{equation*}

The reduction of entanglement with increasing $\gamma $ is expected. The
state $\mathcal{N}(\gamma )e^{-\gamma \hat{J}_{z}}\left\vert
m_{y}=0\right\rangle $ is maximally entangled for $\gamma =0$ and the
prefactor $e^{-\gamma \hat{J}_{z}}$ corresponds to a local non-unitary
operation which always decreases the amount of entanglement. For large
values of $\gamma $ the state becomes eventually separable.

\begin{figure}[tbh]
\begin{center}
\includegraphics[width= 8 true cm]{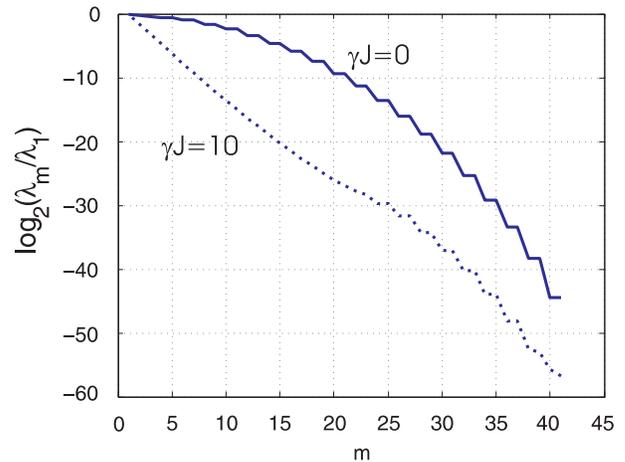}
\end{center}
\caption{Ordered distribution of normalized Schmidt numbers for different
values of $\protect\gamma J$. }
\label{bipartite-entangle-4}
\end{figure}

The most peculiar feature of the von Neumann entropy is the change of the
scaling behavior with $J_1$ from $S\sim \log_2 J_1 $ for $\gamma < \gamma_{%
\mathrm{crit}}$ to $S\sim \log_2 J_1/J$ for $\gamma\ge \gamma_{\mathrm{crit}%
} $. The role of $\gamma_{\mathrm{crit}}=J^{-1}$ and the change of the
scaling behavior is also reflected in the distribution of ordered Schmidt
numbers. As can be seen from fig.\ref{bipartite-entangle-4} the fall-off of
the Schmidt numbers $\lambda_m$ becomes exponential when $\gamma J$ exceeds
unity.

%%%%%%%%%%%%%%%%%%%%%%%%%%%%%%%%%%%%%%%%%%%%%%%%%%%%%%%%%%%%%%%%%%%%%%%%%%%%

%%%%%%%%%%%%%%%%%%%%%%%%%%%%%%%%%%%%%%%%%%%%%%%%%%%%%%%%%%%%%%%%%%%%%%%%%%%%%

There is no obvious distinction of the point $\gamma =J^{-1}$ in the
properties of the system. The system does not undergo a phase transition at
this point. Due to the SUSY the qualitative structure of the spectrum is the
same for all values of $\gamma $ and there is alwas an energy gap between
the ground and first excited state. Thus the question remains whether there
are any physical signatures for the change of the scaling behavior of the
entanglement at $\gamma =\gamma _{\mathrm{crit}}$.

%%%%%%%%%%%%%%%%%%%%%%%%%%%%%%%%%%%%%%%%%%%%%%%%%%%%%%%%%%%%%%%%%%%%%%%%%%%
%%%%%%%%%%%%%%%%%%%%%%%%%%%%%%%%%%%%%%%%%%%%%%%%%%%%%%%%%%%%%%%%%%%%%%%%%%%

\section{Geometric estimate for global entanglement}

%%%%%%%%%%%%%%%%%%%%%%%%%%%%%%%%%%%%%%%%%%%%%%%%%%%%%%%%%%%%%%%%%%%%%%%%%%%
%%%%%%%%%%%%%%%%%%%%%%%%%%%%%%%%%%%%%%%%%%%%%%%%%%%%%%%%%%%%%%%%%%%%%%%%%%%

In the previous section we have discussed the bi-partite entanglement of two
partitions of the $N$ spin $1/2$ system. We have seen (cf. fig\ref%
{bipartite-entangle-3}) that for $\gamma \geq \gamma _{\mathrm{crit}}$ the
von Neumann entropy has a maximum value independent on $J$. In this section
we will quantitatively analyze this maximum by determining the $N$-partite
or global entanglement $E_{N}$ of the SUSY ground state (\ref{SUSY-state})
which is an upper bound to the bipartite entanglement $E_{2}$. Although it
is not possible to obtain an analytic expression for $E_{N}$, we can
determine a very good estimate for it given by the geometric measure of
entanglement.

%%%%%%%%%%%%%%%%%%%%%%%%%%%%%%%%%%%%%%%%%%%%%%%%%%%%%%%%%%%%%%%%%%%%%%%%%%%

\subsection{Relative entropy and geometric measure of entanglement}

%%%%%%%%%%%%%%%%%%%%%%%%%%%%%%%%%%%%%%%%%%%%%%%%%%%%%%%%%%%%%%%%%%%%%%%%%%%

A many-particle state is called $N$-partite separable, if it can be written
as a product of states of all $N$ sub-systems. Obviously a bi-partite
entangled state is always $N$-partite entangled but not vice versa. A
quantitative measure of many-particle or \textit{global} entanglement of a
state $\rho $ is the minimum relative entropy which determines the minimum
distance between $\rho $ and the set $\mathcal{S}_{N}$ of $N$-partite
product states $\sigma $ \cite{Vedral2002} 
\begin{equation}
E_{N}=\underset{\sigma \in \mathcal{S}_{N}}{\min }\mathit{\ }S(\rho ||\sigma
)  \label{entanglementglobal}
\end{equation}%
where 
\begin{equation}
S(\rho ||\sigma )\equiv \text{tr}(\rho \log _{2}\rho -\rho \log _{2}\sigma ),
\end{equation}%
$\sigma \in \mathcal{S}_{N}$ being an $N$-partite separable state 
\begin{equation}
\sigma =\sum_{i=1}^{N}p_{i}\rho _{1}^{i}\otimes \rho _{2}^{i}\otimes
...\otimes \rho _{n}^{i}
\end{equation}%
with $p_{i}>0,$ and $\sum_{i}p_{i}=1$. For the bi-partite case $E_{N}$ is
equivalent to the entanglement of formation \cite{Vedral2002}, which in the
case of pure states is identical to the von Neumann entropy.

Since the set $\mathcal{S}_{N}$ is smaller than $\mathcal{S}_{2}$ for any
partioning, $\mathcal{S}_{N}\subset \mathcal{S}_{2}$, it follows immediately 
\begin{equation*}
E_{N}(\Psi )\geq E_{2}(\Psi ),
\end{equation*}%
i.e. the global entanglement represents an upper bound to the bi-partite
entanglement.

In order to compute $E_{N}$ for any state $\rho $, one has to find its
closest product state $\sigma $. This is in general a quite difficult task
and can be done only in very special cases. There is however a lower bound
to $E_{N}$ which gives a good estimate for the behavior of the global
entanglement. This lower bound is the geometric entanglement $E_{G}(\Psi )$ 
\cite{Wei2003,Wei2004}. 
\begin{equation}
E_{G}(\Psi )\ \equiv -2\log _{2}\Lambda _{\max }(\Psi ),
\label{entanglambda}
\end{equation}%
where 
\begin{equation}
\Lambda _{\max }(\Psi )=\underset{\phi }{\max }\left\vert \left\langle \phi
|\Psi \right\rangle \right\vert  \label{lambda def}
\end{equation}%
is the maximum overlap of $|\Psi \rangle $ with an $N$-partite separable
state $|\phi \rangle $. $E_{G}(\Psi )$ is not an entanglement monotone and
thus in the strict sense not a valid measure of entanglement. It does give
however a close lower bound to $E_{N}$ which for some states such as the
Dicke states is a tight bound, i.e. $E_{G}=E_{N}$ \cite{Wei2003}. The
geometric entanglement can easily be calculated for states which are
permutation invariant, which is the case for the SUSY ground state (\ref%
{SUSY-state}).

%%%%%%%%%%%%%%%%%%%%%%%%%%%%%%%%%%%%%%%%%%%%%%%%%%%%%%%%%%%%%%%%%%%%%%%%%%%%%%

\subsection{Geometric measure of entanglement for the SUSY state}

%%%%%%%%%%%%%%%%%%%%%%%%%%%%%%%%%%%%%%%%%%%%%%%%%%%%%%%%%%%%%%%%%%%%%%%%%%%%%%

To calculate the geometric entanglement $E_{G}$ or equivalently the maximum
overlap $\Lambda _{\mathrm{max}}$ of the SUSY ground state (\ref{SUSY-state}%
) with $N$-partite separable states it is sufficient to construct the most
general $N$-partite separable state which is invariant under permutation of
spins \cite{Wei2003}. This state is given by rotations of the state $%
|m_{z}=-J\rangle $: 
\begin{equation*}
|\phi \rangle =\mathrm{e}^{-i\alpha \hat{J}_{z}}\mathrm{e}^{-i\beta \hat{J}%
_{y}}\mathrm{e}^{-i\xi \hat{J}_{z}}\,|m_{z}=-J\rangle .
\end{equation*}%
Calculating the overlap of $|\phi \rangle $ with (\ref{SUSY-state}) and
maximizing it with respect to the real parameters $\alpha ,\beta $ and $\xi $
leads to 
\begin{equation*}
|\phi \rangle =|m_{z}=-J\rangle .
\end{equation*}%
The corresponding entanglement eigenvalue reads 
\begin{equation}
\Lambda _{\max }(\gamma )=\frac{\sqrt{(2J)!}}{2^{J}J!}\frac{e^{\gamma J}}{%
\sqrt{P_{J}(\cosh 2\gamma )}}.  \label{Lambda_max}
\end{equation}

%********************************************************************

\subsubsection{isotropic spin coupling $\protect\gamma=0$}

%********************************************************************

In the isotropic case (\ref{Lambda_max}) reduces to \cite{Wei2003} 
\begin{eqnarray}
\Lambda_{\max}(\gamma=0)=\frac{\sqrt{(2J)!}}{2^{J}J!}
\end{eqnarray}
and thus the geometric entanglement is given by $E_G(\Psi)=\frac{1}{2}
\log_2 J$. Since the SUSY state for $\gamma=0$ is the Dicke state $%
|J,m_y=0\rangle$ the geometric entanglement is a tight lower bound to the
relative entropy and thus 
\begin{eqnarray}
E_N(\Psi)=E_G(\Psi)=\frac{1}{2}\log_2 J.
\end{eqnarray}

%************************************************************************

\subsubsection{anisotropic spin coupling $\protect\gamma\ne 0$}

%********************************************************************

It is obvious that the largest entanglement is obtained for $\gamma=0$ where
the maximum overlap $\Lambda_{\max}$ with separable states is smallest. On
the other hand for $\gamma \rightarrow \infty$ the state becomes identical
to the separable state $|m_z=-J\rangle$. The same conclusion can of course
be obtained from eq.(\ref{Lambda_max}) employing the asymptotic expansion of
the Legendre polynomials.

%%%%%%%%%%%%%%%%%%%%%%%%%%%%%%%%%%%%%%%%%%%%%%%%%%%%%%%%%%%%%%%%%%%%%%%%%%%%
\begin{figure}[htb]
\begin{center}
\includegraphics[width= 7 true cm]{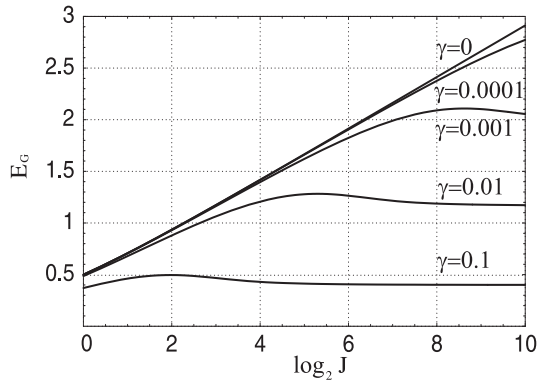}
\end{center}
\caption{Geometric measure of entanglement as function of $\log_2 J$ for
different values of the anisotropy parameter $\protect\gamma$. One
recognizes a saturation at $J \ge \protect\gamma_{\mathrm{crit}}^{-1}$. }
\label{global-entangle}
\end{figure}
%%%%%%%%%%%%%%%%%%%%%%%%%%%%%%%%%%%%%%%%%%%%%%%%%%%%%%%%%%%%%%%%%%%%%%%%%%%%%

In fig.\ref{global-entangle} we have plotted the geometric entanglement as a
function of $J$ for different values of $\gamma $. For sufficiently small
values of $J$ one recognizes a logarithmic growth which saturates when $J$
exceeds the value $\gamma ^{-1}$. One can easily obtain an analytic
expression for the saturation value of $E_{G}$. Making use of the
asymptotics of the Legendre polynomials for large $J$ and $\gamma \neq 0$ 
\begin{equation*}
P_{J}(\cosh 2\gamma )\quad _{\overrightarrow{_{\text{large J}\text{ }}}%
}\quad \frac{1}{\sqrt{1-e^{-4\gamma }}}e^{2\gamma J}\frac{(2J-1)!!}{2^{J}J!}
\end{equation*}%
one arrives at the simple expression 
\begin{equation*}
\Lambda _{\max }(\gamma )=(1-e^{-4\gamma })^{1/4}
\end{equation*}%
leading to 
\begin{equation*}
E_{G}(\gamma ,J)\,\,\overrightarrow{_{\text{large J}\text{ }}}\,\,-\frac{1}{2%
}\log _{2}(1-e^{-4\gamma }).
\end{equation*}%
Comparing the numerical values for $E_{2}$ obtained in the previous section
one finds that $E_{G}<E_{2}$. This shows that in the case of non-isotropic
coupling $\gamma \neq 0$ the geometric entanglement is not a tight lower
bound to the global entanglement, i.e. here $E_{G}<E_{N}$. Thus $E_{G}$ can
only be used as qualitative measure for the global entanglement.

%%%%%%%%%%%%%%%%%%%%%%%%%%%%%%%%%%%%%%%%%%%%%%%%%%%%%%%%%%%%%%%%%%%%%%%%%%%%%%%
%%%%%%%%%%%%%%%%%%%%%%%%%%%%%%%%%%%%%%%%%%%%%%%%%%%%%%%%%%%%%%%%%%%%%%%%%%%%%%%

\section{Conclusions}

%%%%%%%%%%%%%%%%%%%%%%%%%%%%%%%%%%%%%%%%%%%%%%%%%%%%%%%%%%%%%%%%%%%%%%%%%%%%%%%
%%%%%%%%%%%%%%%%%%%%%%%%%%%%%%%%%%%%%%%%%%%%%%%%%%%%%%%%%%%%%%%%%%%%%%%%%%%%%%%

In the present paper we have studied the bi-partite entanglement between
blocks of spins in the anti-ferromagnetic Lipkin-Meshkov-Glick model, which
describes a nonlinear coupling of collective spins, under conditions of
super-symmetry. The super-symmetry of the model allows for an explicit
construction of the ground state which undergoes a smooth transition from a
separable to a maximally entangled state when changing the anisotropy of the
collective spin coupling. Making use of the Clebsch-Gordan decomposition of
angular momenta, the von Neumann entropy which quantifies the bi-partite
entanglement can be calculated analytically in the isotropic case or
numerically in the case of anisotropic coupling. Although the structure of
the spectrum stays always the same with one nondegenerate ground state and
pairwise degenerate excited states, and no level crossing or merging occurs,
the entanglement shows a discontinuous behavior at the isotropy point. When
the anisotropy parameter $\gamma $ vanishes exactly, the von Neumann entropy
grows logarithmically with the number of particles in the subsystem. For any
nonvanishing value of $\gamma $ (in the thermodynamic limit) the entropy
saturates at a finite value determined by $\gamma $. The maximum bi-partite
entanglement can be estimated by the geometric measure of global
entanglement, which has been determined analytically. Furthermore in this
case the entropy becomes a function of the ratio of particle number in the
subsystem to the total particle number rather than a function of the
subsystem size alone. For finite systems the transition between the two
cases happens at a small but finite value of $\gamma $ corresponding to the
inverse of the total number of spins. A discontinuous scaling behavior of
entanglement is usually attributed to level crossings and quantum phase
transitions. This is not the case in the present system and the question
remains whether there are any physical signatures of the discontinuous
transition.

%%%%%%%%%%%%%%%%%%%%%%%%%%%%%%%%%%%%%%%%%%%%%%%%%%%%%%%%%%%%%%%%%%%%%%%%%%%%%%%

\section*{Acknowledgement}

This work was supported by the DFG through the SPP ``Quantum Information''
as well as the European network QUACS. C.I. is supported by a fellowship
from the EU through the Marie-Curie trainingssite at the TU Kaiserslautern.

%%%%%%%%%%%%%%%%%%%%%%%%%%%%%%%%%%%%%%%%%%%%%%%%%%%%%%%%%%%%%%%%%%%%%%%%%%%%%%%
%%%%%%%%%%%%%%%%%%%%%%%%%%%%%%%%%%%%%%%%%%%%%%%%%%%%%%%%%%%%%%%%%%%%%%%%%%%%%%%

%\input{Lipkinbib.bib}

%%%%%%%%%%%%%%%%%%%%%%%%%%%%%%%%%%%%%%%%%%%%%%%%%%%%%%%%%%%%%%%%%%%%%%%%%%%%%%%
%%%%%%%%%%%%%%%%%%%%%%%%%%%%%%%%%%%%%%%%%%%%%%%%%%%%%%%%%%%%%%%%%%%%%%%%%%%%%%%

%%%%%%%%%%%%%%%%%%%%%%%%%%%%%%%%%%%%%%%%%%%%%%%%%%%%%%%%%%%%%%%%%%%%%%%%%%%%%%%
%%%%%%%%%%%%%%%%%%%%%%%%%%%%%%%%%%%%%%%%%%%%%%%%%%%%%%%%%%%%%%%%%%%%%%%%%%%%%%%


\begin{thebibliography}{99}

\bibitem{Schroedinger} E. Schr\"odinger, Naturwissenschaften
{\bf 23}, 807 (1935); {\it ibid} 823 (1935); {\it ibid} 844 (1935).

\bibitem{EPR} A. Einstein, L. Podolski and N. Rosen,
Phys. Rev. {\bf 47}, 777 (1935).

\bibitem{Bell} J. S. Bell, {\it Speakable and Unspeakable in Quantum Mechanics}
(Cambridge University Press, Cambridge, 1987)

\bibitem{Nielsen-Chuang} M. Nielsen and I. Chuang, {\it Quantum Computation and
Quantum Communication} (Cambridge University Press, Cambridge 2000).

\bibitem{Preskill2000} J. Preskill, J. Mod. Opt. {\bf 47}, 127 (2000).

\bibitem{Sachdev} S. Sachdev, {\it Quantum Phase Transitions} (Cambridge Univ. Press, Cambridge, 1999).

\bibitem{Wooters1998} W. K. Wootters,
Phys. Rev. Lett {\bf 80}, 2245 (1998).

\bibitem{OConnor2001} K. M. O'Connor and W.K. Wootters,
Phys. Rev. A {\bf 63}, 052302 (2001).

\bibitem{Arnesen2001} M. C. Arnesen, S. Bose and V. Vedral,
Phys. Rev. Lett. {\bf 87}, 017901 (2001).

\bibitem{Osborne2002} T. J. Osborn and M. A. Nielsen,
Phys. Rev. A {\bf 66}, 032110 (2002).

\bibitem{Osterloh2002} A. Osterloh, L. Amico, G. Falci, and R. Fazio,
Nature {\bf 416}, 608 (2002).

\bibitem{Vidal2003} G. Vidal, J. I. Latorre, E. Rico, and A. Kitaev,
Phys. Rev. Lett. {\bf 90}, 227902 (2003).

\bibitem{Lipkin1965} H. Lipkin, N. Meshkov and A. Glick, Nucl. Phys.
\textbf{62}, 188 (1965).

\bibitem{Witten1981} E. Witten, Nucl. Phys. B \textbf{188}, 513 (1981).

\bibitem{Cooper1995} F. Cooper, A. Khane and U. Sukhamte,
Physics Reports \textbf{251}, 267 (1995).

\bibitem{Unanyan2003} R. G.  Unanyan and M. Fleischhauer,
Phys. Rev. Lett. \textbf{90} 133601 (2003).


\bibitem{Vidal2004b} J. Vidal, R. Mosseri, and J. Dukelsky,
Phys. Rev. A {\bf 69}, 054101 (2004).

\bibitem{Stockton2003} J. K. Stockton, J. M. Geremia, A. C. Doherty and
H. Mabuchi, Phys. Rev. A \textbf{67}, 022112 (2003).

\bibitem{Vidal-preprint} J. I. Latorre, R. Orus, E. Rico and
J. Vidal,  preprint quant-ph/0409611.

\bibitem{Bennett1996} C. H. Bennett et al., Phys. Rev. A
\textbf{53}, 2046 (1996).


\bibitem {Wei2003} Wei and P.M. Goldbart, Phys. Rev. A
\textbf{68},042307 (2003).

\bibitem {Wei2004} T. Wei et al., preprint, quant-ph/0405002.

\bibitem{Vedral2002} V. Vedral, Rev. Mod. Phys. \textbf{74}, 197 (2002).

\bibitem{Peres-book} A. Peres, {\it{Quantum Mechanics: Concepts and Methods}},

(Kluwer, Dordrecht, 1993).


\bibitem{Beckner1997} W. Beckner and M. Pearson, Bull. London Math. Soc.,
\textbf{30}, 80 (1997).

\bibitem {Wigner}E. P. Wigner, \textit{Group Theory} (Academic, New York, 1959).

\bibitem{Biedenharn1981} L.C. Biedenharn and J.D. Louck,
{\it{Angular Momentum in Quantum Physics}} (Addison-Wesley, 1981)

\end{thebibliography}
\end{document}